\documentclass[%
superscriptaddress,
preprint,
 amsmath,amssymb,
 aps,
]{revtex4-1}

\usepackage{graphicx,setspace}% Include figure files
\usepackage{dcolumn}% Align table columns on decimal point
\usepackage{bm}% bold math
\usepackage[mathlines]{lineno}% Enable numbering of text and display math
\usepackage[colorlinks=true,citecolor=teal,linkcolor=olive,anchorcolor=green,urlcolor=teal]{hyperref}
\usepackage{mathrsfs}
\usepackage{xcolor}
\usepackage{float}
\usepackage[normalem]{ulem}
\usepackage[utf8]{inputenc}
\usepackage{tensor}
\usepackage{physics}
\usepackage{blindtext}
\usepackage{slashed}

\newcommand{\be}{\begin{equation}}
\newcommand{\ee}{\end{equation}}
\newcommand{\ba}{\begin{eqnarray}}
\newcommand{\ea}{\end{eqnarray}}

\begin{document}

\title{Holographic complexity of rotating black holes with conical deficits}
\author{Ming Zhang}
\email{mingzhang@jxnu.edu.cn}
\affiliation{Department of Physics, Jiangxi Normal University, Nanchang 330022, China}
\author{Chaoxi Fang}
\email{chaoxi.f@jxnu.edu.cn}
\affiliation{Department of Physics, Jiangxi Normal University, Nanchang 330022, China}
\author{Jie Jiang}
\email{jiejiang@mail.bnu.edu.cn (corresponding author)}
\affiliation{College of Education for the Future, Beijing Normal University, Zhuhai 519087, China}

\date{\today}

\begin{abstract}
Based on the complexity equals action (CA) and complexity equals volume (CV) conjectures, we investigate the holographic complexity of a slowly accelerating Kerr-AdS black hole in the bulk Einstein gravity theory which is dual to holographic states with rotation and conical deficits in the boundary quantum system. Upon obtaining an implicit form of the Wheeler-DeWitt patch, we evaluate the action and show that the growth rate of the CA complexity violates volume-scaling formulation in large black hole limit due to the non-trivial contribution from the not-too-small acceleration of the black hole. Moreover, in an ensemble with fixed entropy, pressure, and angular momentum, we also find that complexity of formation decreases with both the average and difference of the conical deficits on the poles when the black hole is close to the static limit but increases with the deficits when the black hole is close to the extremal regime. 
\end{abstract}
\maketitle

\section{Introduction}
As the slogan ``entanglement is not enough'' \cite{Aaronson:2016vto} claims, the quantum computational complexity of conformal field theory (CFT) quantum gravity states can be a new measure of information of spacetime \cite{Susskind:2014moa}. Among many bulk observable proposals of the boundary CFT states' complexity, the complexity $=$ action $(\mathrm{CA})$ \cite{Brown:2015bva,Brown:2015lvg} and complexity $=$ volume $(\mathrm{CV})$ \cite{Susskind:2014rva,Stanford:2014jda} formulations are two quintessential ones. The former proposes that the complexity relates to gravitational action evaluated on the Wheeler-DeWitt (WDW) patch of the spacetime, which is anchored on boundary time slice $\Sigma$, i.e., $C_{\mathrm{A}}=I_{\mathrm{WDW}}/(\pi\hbar),$ where the gravitational action $I_{\mathrm{WDW}}$ is explicitly given in \cite{Lehner:2016vdi}. The latter one implies that $C_{\mathrm{\mathcal{V}}}=\max _{\partial \mathcal{B}=\Sigma}\left[\mathcal{V}(\mathcal{B})/(G_{N} \ell_{\mathrm{bulk}})\right]$, which means that the complexity of the thermofield double (TFD) state at the boundary section $\Sigma$ can be computed through evaluating the volume of a codimension-one maximal hypersurface $\mathcal{B}$ in bulk, where $G_N$ is the Newton's constant in the bulk gravitational theory and the length scale ${\ell}_{\mathrm{bulk}}$ can be taken to be the Anti-de Sitter (AdS) radius $\ell$ \cite{Brown:2015bva,Couch:2018phr}. 

There is dramatic progress toward investigating and checking the CA and CV conjectures by studying both holographic complexity \cite{Susskind:2014jwa,Chapman:2016hwi,Carmi:2016wjl,Moosa:2017yvt,Cai:2017sjv,Brown:2017jil,Carmi:2017jqz,An:2018xhv,Chapman:2018dem,Chapman:2018lsv,Goto:2018iay,Flory:2018akz,Bernamonti:2019zyy,Bernamonti:2020bcf,Barbon:2019tuq,Iliesiu:2021ari,Auzzi:2019vyh} and circuit complexity in quantum field theory \cite{Jefferson:2017sdb,Chapman:2018hou,Caputa:2017urj,Bhattacharyya:2018wym,Caputa:2018kdj,Chagnet:2021uvi,Flory:2020eot}. Most of the researches focused on highly symmetric cases, except those including local quenches \cite{Ageev:2019fxn,DiGiulio:2021noo} and with defects \cite{Chapman:2018bqj,Braccia:2019xxi}. Efforts to estimate late time growth rate of the holographic complexity for rotating black holes can be seen in \cite{Brown:2015lvg,Lehner:2016vdi,Cai:2016xho,Auzzi:2018zdu,Auzzi:2018pbc,Frassino:2019fgr}. For the rotating black holes with spacetime dimensions greater than three, one main obstacle is explicitly obtaining closed forms of the WDW patch. As a remarkable progress, recently the holographic complexity of the odd-dimensional Myers-Perry black holes with equal orthogonal angular momenta was calculated in \cite{AlBalushi:2020ely,AlBalushi:2020rqe} due to symmetry enhancement. It is till the recent time that the null hypersurface foliations of the Kerr-(A)dS \cite{AlBalushi:2019obu}, Kerr-Newman-AdS \cite{Imseis:2020vsw} were analyzed and implicit descriptions were given. Then partial progress was made in \cite{Bernamonti:2021jyu} to precisely calculate the growth rate of complexity at late times by taking all necessary terms defining the action into account and to analyze the complexity of formation within the framework of the CV proposal. Being consistent with the results in \cite{AlBalushi:2020ely,AlBalushi:2020rqe}, it was found that the relation $\lim _{t_{b} \rightarrow \infty} d C/d t_{b} \propto P \Delta V$ is also respected at the leading order at late times, where $P$ is the thermodynamic pressure and $\Delta V$ is the difference between thermodynamic volumes defined on the event horizon and Cauchy horizon. 

The goal of this paper is to investigate the holographic complexity of rotating black holes with conical deficits. Specifically, we study the slowly accelerating Kerr-AdS black hole in Einstein gravity described by C-metric \cite{Plebanski:1976gy}. The motivations of our work are as follows. The branch of accelerating black holes amongst the panoply of the black hole family is less known. These objects are characterized by the conical singularities on the poles, which induce acceleration. We wonder how the complexities encode information about the conical deficits in the boundary quantum system. Indeed, it was shown in \cite{Appels:2016uha,Anabalon:2018ydc,Anabalon:2018qfv,Gregory:2019dtq} that the conical deficits can be viewed as charges of the black hole. Thus, we will study how the additional charges affect the complexity of the black hole. Moreover, astrophysical black holes on all scales, from stellar binaries to galaxies and active galactic nuclei, are expected to be rotating \cite{Kleihaus:2011tg}; on the other hand, accelerating supermassive black holes connected to cosmic strings which as topological conical defects pulling on the black holes \cite{Appels:2016uha} could reside in the centers of galaxies, and the velocities of them must be small ($\lesssim 100 \mathrm{~km} / \mathrm{s}$) \cite{morris2017nonthermal,Vilenkin:2018zol,Ashoorioon:2022zgu}. Besides, the setup of slow acceleration admits equilibrium thermodynamics for the system, and the application of the holographic principle \cite{Maldacena:1997re,Lu:2008jk,Gao:2019vcc} between the astrophysical black hole and the boundary CFT is unambiguous and straightforward \cite{Anabalon:2018ydc}.  As pointed out in \cite{Bernamonti:2021jyu}, in the four-dimensional case, limited symmetry is endowed, and it is impossible to evaluate the full-time evolution of the complexity. Thus we will focus on the CA complexity at late times and the CV complexity of formation.

We arrange the rest of the paper as follows. In the following section, we will give a brief review of the accelerating Kerr-AdS black hole. In Sec. \ref{wdwpat}, we will formulate the WDW patch for the rotating black hole with a conformal factor. In Sec. \ref{cacomp}, we will evaluate the action of the black hole characterized by conical deficits and calculate the complexity growth rate at late times. In Sec. \ref{cvfor}, we will investigate the CV complexity of formation for the black hole. The last section is devoted to our conclusions. We set $\hbar=1$ throughout the paper.

\section{Accelerating Kerr-AdS black hole}\label{accbh}
The metric of the accelerating Kerr-AdS black hole is
\begin{equation}\label{met}
\begin{aligned}
\tilde{g}_{\mu \nu}=&d \tilde{s}^{2}=\frac{1}{\Upsilon^{2}}\left[-\frac{\Delta_{r}}{\rho^{2}}\left(\frac{d t}{\alpha}-a \sin ^{2} \theta \frac{d \phi}{K}\right)^{2}\right. \\
&\left.+\rho^{2}\left(\frac{d r^{2}}{\Delta_{r}}+\frac{d \theta^{2}}{\Delta_{\theta}}\right)+\frac{\Delta_{\theta} \sin ^{2} \theta}{\rho^{2}}\left[\frac{a d t}{\alpha}-\left(r^{2}+a^{2}\right) \frac{d \phi}{K}\right]^{2}\right],
\end{aligned}
\end{equation}
where
\begin{equation*}
\begin{aligned}
\Upsilon &=1+A r \cos \theta, \;\rho^{2} =r^{2}+a^{2} \cos ^{2} \theta,\\
\Delta_{r} &=\left(1-A^{2} r^{2}\right)\left(r^{2}-2 m r+a^{2}\right)+\frac{\left(r^{2}+a^{2}\right) r^{2}}{\ell^{2}} ,\\
\Delta_{\theta} &=1+2 m A \cos \theta+\left(a^{2} A^{2}-\frac{a^{2}}{\ell^{2}}\right) \cos ^{2} \theta.
\end{aligned}
\end{equation*}
$m, a$ are mass parameter and angular momentum per unit mass of the black hole, $\ell$ is AdS radius, $A$ reflects the acceleration of the black hole, the normalization factor $\alpha$ is proposed to give correct time coordinate $\tau=\alpha t$ corresponding to the time of an asymptotic observer. The conformal factor $\Upsilon$ determines a conformal boundary by $\Upsilon (r_b)=0$. The Cauchy horizon $r_-$ and the event horizon $r_+$ of the black hole are determined by $\Delta_r(r_\pm)=0$. There are conical singularities located on the two poles of the black hole, which are induced by conical deficit angles that are controlled by the parameter $K$.

The thermodynamic first law of the black hole is
\begin{equation}\label{tfl}
\delta M=T \delta S+\Omega \delta J-\lambda_{+} \delta \mu_{+}-\lambda_{-} \delta \mu_{-}+V \delta P,
\end{equation}
where $M, T, S, J, \Omega, P, V, \mu_{\pm}, \lambda_\pm$ are individually the mass, temperature, entropy, angular momentum, angular velocity, pressure, thermodynamic volume, tensions, and thermodynamic length of the black hole \cite{Anabalon:2018qfv}. The integral Smarr relation is
\begin{equation}\label{sma}
M=2(T S+\Omega J-P V)+\Phi Q,
\end{equation}
which is free of the tensions $\mu_\pm$ due to their scaling properties and a proper time scaling parameter $\alpha=\sqrt{\left(\Xi+a^{2} / \ell^{2}\right)\left(1-A^{2} \ell^{2} \Xi\right)}/(1+a^{2} A^{2})$
should be chosen, where $\Xi=1-a^{2}/\ell^{2}+A^{2}a^{2}$.
The tensions on the poles originate from the conical deficits in the relation
\begin{equation}\label{mupm}
\mu_{\pm}=\frac{1}{4}\left(1-\frac{\Delta_{\theta\pm}}{K}\right),
\end{equation}
where $\Delta_{\theta_+}\equiv \Delta_\theta (\theta=0)$ and $\Delta_{\theta_-}\equiv \Delta_\theta (\theta=\pi)$.
The thermodynamic quantities of the black hole are \cite{Anabalon:2018qfv}
\begin{equation}
\begin{aligned}
M &=\frac{m\left(\Xi+a^{2} / \ell^{2}\right)\left(1-A^{2} \ell^{2} \Xi\right)}{K \Xi \alpha\left(1+a^{2} A^{2}\right)},\; T =\frac{f_{+}^{\prime} r_{+}^{2}}{4 \pi \alpha\left(r_{+}^{2}+a^{2}\right)}, \\ S&=\frac{\pi\left(r_{+}^{2}+a^{2}\right)}{K\left(1-A^{2} r_{+}^{2}\right)},\; J=\frac{m a}{K^{2}}, \; P=\frac{3}{8 \pi \ell^{2}},\\\Omega&=\frac{K a}{\alpha\left(r_{+}^{2}+a^{2}\right)}+\frac{a K\left(1-A^{2} \ell^{2} \Xi\right)}{\ell^{2} \Xi \alpha\left(1+a^{2} A^{2}\right)}.
\end{aligned}
\end{equation}

To ensure that the metric (\ref{met}) describes a single slowly accelerating black hole suspended by cosmic strings in asymptotically AdS space, several constraints should be imposed on the parameters: $\Delta_\theta >0$ for $\theta\in [0, \pi]$; $\Delta_r(r_b) =0$ does not have any real roots; $\Delta_r(r)=0$ has not less than one root for $A r\in (0, 1)$ \cite{Abbasvandi:2019vfz}; $\ell \Omega<1$ \cite{Bernamonti:2021jyu}, which is equivalent to that the black hole keeps globally timelike Killing vector outside the event horizon and is in thermodynamical equilibrium all the way to conformal infinity with rotating thermal radiation \cite{Hawking:1998kw}.

\section{Forms of the WDW patch}\label{wdwpat}
The null hypersurfaces that surround the codimension-0 WDW patch are given by $g^{\alpha \beta} \partial_{\alpha} \Phi \partial_{\beta} \Phi=0$, with $\Phi(x)=$ const. a constraint equation. By defining the ingoing and outgoing Eddington-Finkelstein coordinates 
\begin{equation}
\begin{aligned}
v=t+r_{*}(r,\theta),\; u=t-r_{*}(r,\theta),
\end{aligned}
\end{equation}
with $r_{*}=r_{*}(r, \theta)$ the implicit tortoise coordinate, we can get null hypersurfaces $v=\text{const}.$ and $u=\text{const}.$ that penetrate the horizon of a conformal metric
\begin{equation}
g_{\mu \nu}=\Upsilon^{2} \tilde{g}_{\mu \nu}.
\end{equation}
This conformal transformation keeps the light cones of the spacetime invariant.
Then the ingoing and outgoing null generators of the null hypersurfaces for the accelerating Kerr-AdS black hole are determined by $v=$ const. and $u=$ const.. Specifically, we have 
\begin{equation}
\Delta_{r}\left(\partial_{r} r_{*}\right)^{2}+H_{\theta}\left(\partial_{\theta} r_{*}\right)^{2}=\alpha^{2}\left[\frac{\left(r^{2}+a^{2}\right)^{2}}{\Delta_{r}}-\frac{a^{2} \sin ^{2} \theta}{H_{\theta}}\right]
\end{equation}
for the null hypersurface equation.

By using a function $\lambda=\lambda(r, \theta)$, we can separate the equation into one with radial coordinate and the other one with longitudinal coordinate \cite{AlBalushi:2019obu},
\begin{equation}
\begin{aligned}\partial_{r} r_{*}=\frac{Q}{\Delta_{r}}, \;
\partial_{\theta} r_{*}=\frac{\mathcal{P}}{\Delta_{\theta}},
\end{aligned}
\end{equation}
where
\begin{equation}
\begin{aligned}
&Q^{2}(r)=\alpha^{2}\left[\left(r^{2}+a^{2}\right)^{2}-a^{2} \lambda \Delta_{r}\right], \\
&\mathcal{P}^{2}(\theta)=\alpha^{2} a^{2}\left[\lambda \Delta_{\theta}-\sin ^{2} \theta\right].
\end{aligned}
\end{equation}

\begin{figure*}[t!]
\vspace{-5mm}
\begin{center}
\includegraphics[width=3in,angle=0]{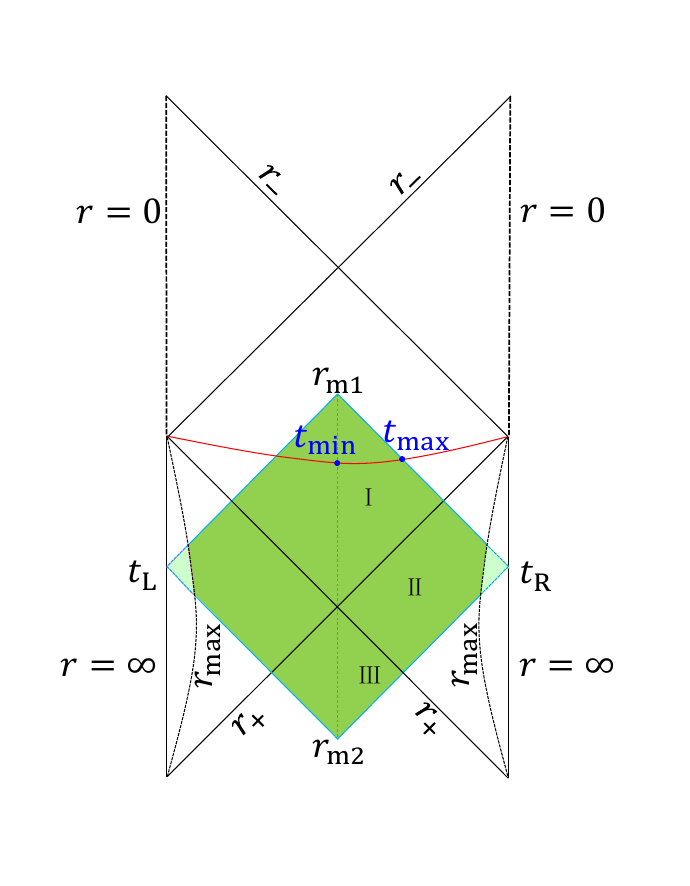}
\includegraphics[width=3in,angle=0]{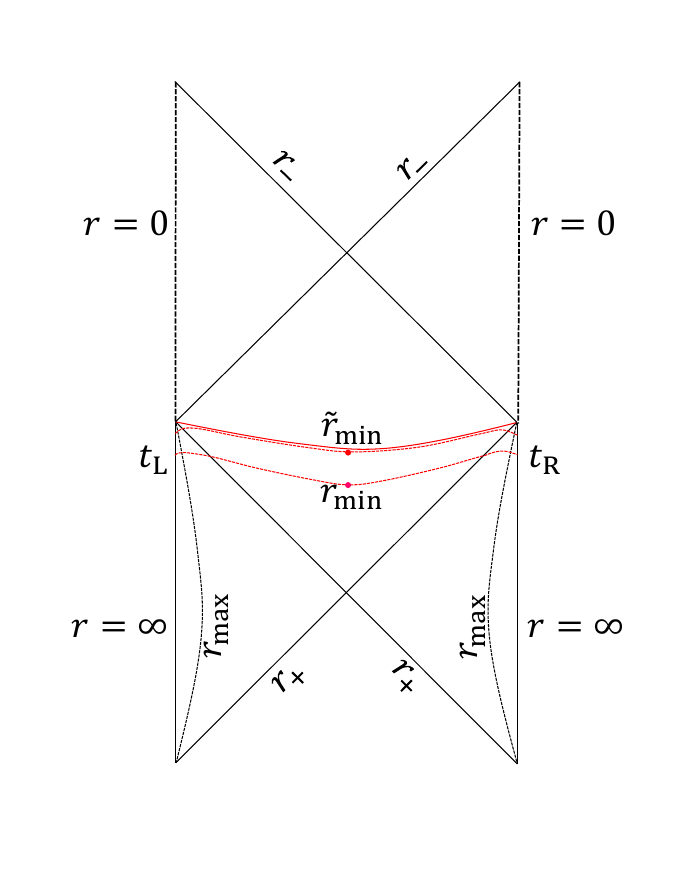}
\end{center}
\vspace{-13mm}
 \caption {A Penrose diagram of accelerating Kerr-AdS black hole. Left panel: the shaded green region bounded by four null sheets represents the WDW patch of boundary time slices $t_\textrm{L}=t_\textrm{R}$. Right panel: maximal slices of the CV conjecture are denoted as $r_\textrm{min}$ which tends to $\tilde{r}_\textrm{min}$ at late times. }\label{pic0}
\end{figure*}

In fact, viewing $\lambda$ as a variable, we have
\begin{equation}
d r_{*}(r, \theta, \lambda)=\frac{Q}{\Delta_{r}} d r+\frac{\mathcal{P}}{H_{\theta}} d \theta+\frac{a^{2}}{2} \mathcal{F} d \lambda
\end{equation}
for the tortoise coordinate $r_*$, where
\begin{equation}
\mathcal{F}(r, \theta, \lambda)=\int_{r}^{\infty} \frac{d r^{\prime}}{Q\left(r^{\prime}, \lambda\right)}+\int_{0}^{\theta} \frac{d \theta^{\prime}}{\mathcal{P}\left(\theta^{\prime}, \lambda\right)}+g^{\prime}(\lambda)=0,
\end{equation}
with $g[\lambda(r, \theta)]$ some specific function and $g^\prime (\lambda)= dg/d\lambda$.
Then due to $d \mathcal{F}=0$, we have $\mu d \lambda=-d r/Q+d \theta/\mathcal{P}$, where $\mu=\mu(r, \theta)\equiv-\partial_{\lambda} \mathcal{F}$. In what follows, we will see that to achieve our goal, explicit solutions of the null hypersurface equations and the related induced metric are not necessary. 

We pay attention to the WDW patch's future part and set the null normal one-forms of the null hypersurfaces given by $v=\text{const}.$ and $u=\text{const}.$ to be outward directed from the boundary of the WDW patch, 
\begin{equation}
k_{R \mu}=\partial_{\mu} v=\left(1, \frac{Q}{\Delta_r}, \frac{\mathcal{P}}{\Delta_{\theta}}, 0\right),\;
k_{L \mu}=-\partial_{\mu} u=\left(-1, \frac{Q}{\Delta_r}, \frac{\mathcal{P}}{\Delta_{\theta}}, 0\right)\label{klmu}.
\end{equation}
They define an affine parametrization of the WDW patch's null direction through the null generator $\lambda $ by $\partial_{\lambda} \equiv k^{\mu} \partial_{\mu}$.

Using the intrinsic coordinates $y^{\mathcal{A}}$ which is spacelike on the WDW patch's boundary, we can define spacelike vectors $e_{\mathcal{A}}^{\mu}=\partial x^{\mu}/\partial y^{\mathcal{A}}$.
First, observing that $k^{\mu} \partial_{\mu} \lambda=0$,
we get $e_{\lambda}^{\mu}=\mu\left(0,-\mathcal{P}^{2} Q \Delta_r/\Sigma^{2}, \mathcal{P} Q^{2} \Delta_{\theta}/\Sigma^{2}, 0\right).$
As the black hole is axially symmetric, we choose the other intrinsic coordinate to be $e_{\phi}^{\mu}=(0,0,0,1)$.
Thus we can obtain the transverse metric $\gamma_{\mathcal{A}\mathcal{B}}$ related to the WDW patch,
\begin{equation}
\gamma_{\mathcal{A}\mathcal{B}}=g_{\mu \nu} e_{\mathcal{A}}^{\mu} e_{\mathcal{B}}^{\nu}=\frac{1}{\Upsilon^2}\left(\begin{array}{cc}
\frac{\alpha^2\mu^{2} \rho^{2} \mathcal{P}^{2} Q^{2}}{\Sigma^{2}} & 0 \\
0 & \frac{\Sigma^{2} \sin ^{2} \theta}{K^2 \rho^{2}}
\end{array}\right),
\end{equation}
whose determinant is $\sqrt{\gamma}=\alpha\mu \mathcal{P} Q \sin \theta/(\Upsilon^2 K)$.

\section{Growth rate of CA complexity}\label{cacomp}
We now investigate the CA proposal, for which we need to evaluate the action composed of the Einstein-Hilbert action, the Gibbons-Hawking-York surface term, the additional surface terms dealing with null segments of boundary, and the counterterm. Schematically, using the conventions in \cite{Carmi:2017jqz}, the action is given by
\begin{equation}\label{ttac}
\begin{aligned}
I_{\mathrm{WDW}}=& I_{\mathrm{bulk}}+I_{\mathrm{GHY}}+I_{\mathrm{joints}}+I_{\kappa}+I_{\mathrm{ct}} \\
=& \frac{1}{16 \pi G_{N}} \int d^{d+1} x \sqrt{|g|}\left[R+\frac{d(d-1)}{\ell^{2}}\right]\\&+\frac{1}{8 \pi G_{N}} \int_{\text {regulator }} d^{d} y \sqrt{|h|} \mathcal{K} \\
&+\frac{1}{8 \pi G_{N}} \int_{\text {joints }} d^{d-1} y \sqrt{\sigma} \mathrm{a}_{\text {joint }}\\&+\frac{1}{8 \pi G_{N}} \int_{\partial \mathrm{WDW}} d \lambda d^{d-1} y \sqrt{\gamma} \kappa \\
&+\frac{1}{8 \pi G_{N}} \int_{\partial \mathrm{WDW}} d \lambda d^{d-1} y \sqrt{\gamma} \Theta \log \left(L_{\mathrm{ct}} \Theta\right) ,
\end{aligned}
\end{equation}
where $\mathcal{K}$ is the trace of the extrinsic curvature of boundary, $\mathrm{a}_{\text {joint }}$ denotes normal on the intersecting segment of the null boundary, $\kappa$ vanishes as the null boundaries of the WDW patch we choose are affinely parameterized, $\Theta$ denotes the expansion. $L_{\textrm{ct}}$ is used to keep reparameterization invariance of the action.

We show the Penrose diagram for the black hole with a WDW patch included in Fig. \ref{pic0}. A symmetric time evolution is chosen, so that we set $t_{\textrm{L}}=t_{\textrm{R}}=t_{\textrm{b}}/2$, where the boundary time $t_\textrm{b}>0$ is defined at $r=\infty$. The future and past tips of the WDW patch are individually denoted as $r_\textrm{m1}$ and $r_\textrm{m2}$. The null-null joints of the WDW patch are $\theta$-dependent, so their radial location can be written as
\begin{align}
\frac{t_{b}}{2}+r^{*}(\infty, \theta)&=r^{*}\left(r_{m 1}, \theta\right), \\ \frac{t_{b}}{2}-r^{*}(\infty, \theta)&=-r^{*}\left(r_{m 2}, \theta\right),
\end{align}
leading to
\begin{align}
\frac{\partial r_{m 1}}{\partial t_{b}} =\left.\frac{\Delta_r}{2 Q}\right|_{r_{m 1}},\;
\frac{\partial r_{m 2}}{\partial t_{b}} =-\left.\frac{\Delta_r}{2 Q}\right|_{r_{m 2}}.\label{rm1tb}
\end{align}
which vanish at late times as $r_{m 1} \rightarrow r_{-},\, r_{m 2} \rightarrow r_{+}$.

For the bulk term, we obtain
\begin{equation}
\begin{aligned}
I_{\mathrm{bulk}}=-\frac{3}{4 \alpha G_{N} \ell^{2} K} \int d \theta d r d t \frac{\rho^{2} \sin \theta}{\Upsilon^4}
\end{aligned}
\end{equation}
 by direct calculation, where we have used on-shell condition $R=-12/\ell^2$. As the WDW patch is left-right symmetric, we can split each branch into three parts $\mathrm{I}, \mathrm{II}$, and $\mathrm{III}$, cf. Fig. \ref{pic0}. By symmetry, $t_{\min }=0$, and as located on the same $v$, we get $t_{\max}+r_{*}(r,\theta)=t_\textrm{b}+r_{*}(\infty)$. Consequently,
\begin{equation}
I_{\mathrm{bulk}}=2\left(I_{\mathrm{bulk}}^{\mathrm{I}}+I_{\mathrm{bulk}}^{\mathrm{II}}+I_{\mathrm{bulk}}^{\mathrm{III}}\right),
\end{equation}
where
\begin{align} % requires amsmath; align* for no eq. number
 I_{\mathrm{bulk}}^{\mathrm{I}}=&-\frac{3}{4\alpha G_{N} \ell^{2} K} \int_{0}^{\pi} d \theta \sin \theta \nonumber \\&\times \int_{r_{m 1}}^{r_{+}} d r \frac{r^{2}+a^{2} \cos ^{2} \theta}{\Upsilon^4}\left[\frac{t_{b}}{2}+r^{*}(\infty, \theta)-r^{*}(r, \theta)\right],\\
 I_{\text {bulk }}^{\mathrm{II}}=&\frac{3}{2\alpha G_{N} \ell^{2} K} \int_{0}^{\pi} d \theta \sin \theta \nonumber \\&\times \int_{r_{+}}^{r_{\max }} d r \frac{r^{2}+a^{2} \cos ^{2} \theta}{\Upsilon^4}\left[r^{*}(r, \theta)-r^{*}(\infty, \theta)\right],\\
 I_{\mathrm{bulk}}^{\mathrm{III}}=&\frac{3}{4\alpha G_{N} \ell^{2} K} \int_{0}^{\pi} d \theta \sin \theta \nonumber \\&\times \int_{r_{m 2}}^{r_{+}} d r \frac{r^{2}+a^{2} \cos ^{2} \theta}{\Upsilon^4}\left[\frac{t_{b}}{2}-r^{*}(\infty, \theta)+r^{*}(r, \theta)\right],
\end{align}
where $r_{\textrm{max}}$ is a $\theta$-dependent but $t$-independent UV cutoff which makes the bulk integral regularized. Then the growth rate of the bulk action is given by
\begin{equation}
\frac{d I_{\mathrm{bulk}}}{d t_{b}}=\frac{3}{4 \alpha K G_{N} \ell^{2}}\left(\left. \int_0^\pi d \theta \sin \theta \mathcal{X}\right|_{r_{m 1}}-\left. \int_0^\pi d \theta \sin \theta \mathcal{X}\right|_{r_{m 2}}\right),
\end{equation}
where $\mathcal{X}=\int d r \rho^{2}/\Upsilon^{4}$.
At late times, it becomes
\begin{equation}\label{cabulk}
\begin{aligned}
\lim _{t_{b} \rightarrow \infty} \frac{d I_{\mathrm{bulk}}}{d t_{b}}=\frac{1}{2G_N K \ell^2\alpha}\left[\frac{r_{-}\left(a^{2}+r_{-}^{2}\right)}{\left(1-A^{2} r_{-}^{2}\right)^{2}}-\frac{r_{+}\left(a^{2}+r_{+}^{2}\right)}{\left(1-A^{2} r_{+}^{2}\right)^{2}}\right].
\end{aligned}
\end{equation}

Now we evaluate the contribution from the boundary terms. Except for the timelike cutoff surface $r=r_{\textrm{max}}$, there are other boundaries near the two poles $\theta=0,\,\pi$, which yield
\begin{equation}
\begin{aligned}
I_{\text {GHY}}&=\frac{1}{8 \pi G_{N}}\left( \int_{r=r_{\max }} d^{3} y \sqrt{|h_r|} \mathcal{K}_r+ \int_{\theta=\epsilon,\,\pi-\epsilon} d^{3} y \sqrt{|h_\theta|} \mathcal{K}_\theta\right),\label{eqregu}
\end{aligned}
\end{equation}
where $\epsilon\ll 1$, $\mathcal{K}\equiv h^{a b} \mathcal{K}_{a b}$ with $\mathcal{K}_{a b}=\frac{\partial x^{\mu}}{\partial y^{a}} \frac{\partial x^{\nu}}{\partial y^{b}} \nabla_{\mu} n_{\nu}$, and $n_\mu$ is the outward normal covector of the hypersurface. The contribution from the $r=r_{\max}$ surface is time-independent, so we directly get $d I_{r=r_{\max}}/d t_{b}=0.$ The contribution from the boundary surfaces $\theta=\epsilon,\,\pi-\epsilon$ is
\begin{equation}
\begin{aligned}
I_{\theta=\epsilon,\,\pi-\epsilon}=&\frac{\delta t}{4 G_N} \left(\int_{r_{m 2}}^{r_{m 1}} \sqrt{\left.-h\right|_{\theta=\varepsilon}} \mathcal{K}_\theta d r+ \int_{r_{m 2}}^{r_{m 1}} \sqrt{\left.-h\right|_{\theta=\pi-\varepsilon}} \mathcal{K}_\theta d r \right)\\=&\frac{\delta t}{4 G_N}\left(X_1 \int_{r_{m 2}}^{r_{m 1}} \frac{1}{(1+Ar)^2} d r +X_2 \int_{r_{m 2}}^{r_{m 1}} \frac{1}{(1-Ar)^2} d r\right),
\end{aligned}
\end{equation}
where we have denoted 
\begin{equation}\label{x1x2}
X_1=\frac{a^2 \left(A^2 \ell^2-1\right)+\ell^2 (2 A m+1)}{\alpha K \ell^2 },\; X_2=-X_1+\frac{4 A m}{\alpha K}.
\end{equation}
Consequently, at late times, we have the growth rate of the boundary term
\begin{equation}\label{caco}
\begin{aligned}
\lim_{t_{b} \rightarrow \infty} \frac{d I_{\theta}}{d t_{b}}=\frac{1}{4G_N}\left[\frac{X_{1}\left(r_{-}-r_{+}\right)}{\left(1+A r_{-}\right) \left(1+A r_{+}\right)}+\frac{X_{2}\left(r_{-}-r_{+}\right)}{\left(1-A r_{-}\right) \left(1-A r_{+}\right)}\right].
\end{aligned}
\end{equation}

In what follows we consider the joint terms. The first kind of joint terms is time-null joints at the intersection of the UV cutoff surface $r=r_{\text{max}}$ and the WDW patch. However, it is evident that the action is independent of the time, so straightforwardly we have
\begin{equation}
\frac{d I_{\text {joints }}^{\text {Time-Null }}}{d t_{b}}=0.
\end{equation}
Besides, using the null normal one-forms for the hypersurfaces $v=$ const. and $u=$ const. in Eq. (\ref{klmu}), we have $k_{L} \cdot k_{R}=2\Upsilon^2 /N^{2}$
where $N^{2}=\rho^{2} \Delta_r \Delta_{\theta}/\Sigma^{2}$ with $\Sigma^{2}=\alpha^2\left[\left(r^{2}+a^{2}\right)^{2} \Delta_{\theta}-a^{2} \Delta_r \sin ^{2} \theta\right]$.
 Whence we get
 \begin{equation}
 \begin{aligned}
\frac{d I_{\text {joints }}^{\text {Null-Null }}}{d t_{b}}&=\left.\frac{1}{4 G_{N}}\left[\frac{\partial}{\partial t_{b}} \int d \lambda \sqrt{\gamma} \log \left(-\frac{\Upsilon^2}{N^{2}}\right)\right]\right|_{r=r_{m 1}}\\&\quad+\left.\frac{1}{4 G_{N}}\left[\frac{\partial}{\partial t_{b}} \int d \lambda \sqrt{\gamma} \log \left(-\frac{\Upsilon^2}{N^{2}}\right)\right]\right|_{r=r_{m 2}}
\end{aligned}
\end{equation}
for the joints at the WDW patch's top and bottom corners.
By employing Eq. (\ref{rm1tb}), we can derive that at late times
\begin{equation}\label{cajoint}
\begin{aligned}
\lim _{t_{b} \rightarrow \infty} \frac{d I_{\text {joints }}^{\text {Null-Null }}}{d t_{b}}=\frac{-\alpha \Delta^{\prime}(r_-)}{4 G_N K\left(1-A^{2} r_{-}^{2}\right)}+\frac{\alpha \Delta^{\prime}(r_+)}{4 G_N K\left(1-A^{2} r_{+}^{2}\right)}.
\end{aligned}
\end{equation}

Lastly, we calculate the counterterm action in Eq. (\ref{ttac}). By definition, we have the expansion $\Theta\equiv\partial_{\zeta} \log \sqrt{\gamma}$, where $\zeta$ is a null coordinate associated with $\tilde{N}^{\mu} k_{\mu}=-1$ with $\tilde{N}^{\mu}$ some null vector \cite{Bernamonti:2021jyu}.  For the future boundary of the WDW patch, we have
\begin{equation}
\begin{aligned}
I_{\mathrm{ct}}&=\left.\frac{1}{4 G_{N}} \int d \theta d r^{*} \frac{\alpha \Delta_r \rho^{2} \sin \theta}{\Upsilon^2 K Q} \Theta \log \left(L_{\mathrm{ct}} \Theta\right)\right|_{r=r_{m1}}\\&\quad +\left.\frac{1}{4 G_{N}} \int d \theta d r^{*} \frac{\alpha \Delta_r \rho^{2} \sin \theta}{\Upsilon^2 K Q} \Theta \log \left(L_{\mathrm{ct}} \Theta\right)\right|_{r=r_{m2}},
\end{aligned}
\end{equation}
which at late times reduces to
\begin{equation}\label{cacount}
\lim _{t_{b} \rightarrow \infty} \frac{d I_{\mathrm{ct}}}{d t_{b}}=0.
\end{equation}

Combining the above results in Eqs. (\ref{cabulk}), (\ref{caco}), (\ref{cajoint}), and (\ref{cacount}), we obtain the CA growth rate of the accelerating Kerr-AdS black hole as
\begin{equation}\label{ltbca}
\begin{aligned}
\lim _{t_{b} \rightarrow \infty} \pi \frac{d C_{\mathrm{A}}}{d t_{b}}&=\frac{r_{+}^{3}-r_{-}^{3}+\ell^{2}\left(r_{+}-r_{-}\right)}{2 G_{N}\left(\ell^{2}-a^{2}\right)}\\&=\left(M-\Omega_{+} J\right)-\left(M-\Omega_{-} J\right)\\&\quad+\frac{X_1 (r_--r_+)}{4G_N(1+Ar_-)(1+Ar_+)}\\&\quad+\frac{X_2 (r_--r_+)}{4G_N(1-Ar_-)(1-Ar_+)}.
\end{aligned}
\end{equation}
It is clear that when $A\to 0$, the CA growth rate reduces to the Kerr case shown in \cite{Bernamonti:2021jyu}.
Using the Smarr relation Eq. (\ref{sma}), Eq. (\ref{ltbca}) can be further reduced to
\begin{equation}
\begin{aligned}
&\lim _{t_{b} \rightarrow \infty} \pi \frac{d C_{\mathrm{A}}}{d t_{b}}=T_{+} S_{+}-T_{-} S_{-}-P\left(V_{+}-V_{-}\right)\\&\quad-\frac{X_1 (r_+-r_-)}{4G_N(1+Ar_-)(1+Ar_+)}-\frac{X_2 (r_+-r_-)}{4G_N(1-Ar_-)(1-Ar_+)},
\end{aligned}
\end{equation}
where $X_1$ and $X_2$ are defined in Eq. (\ref{x1x2}). 

In the large black hole limit, considering that $r_-\ll r_+$, $\ell\ll r_+$, $A \ell\ll 1$, after defining dimensionless quantities $\varpi=\ell/r_+, \,\varrho=r_-/r_+\,,\tilde{A}=A \ell$, we further obtain
\begin{equation}
\begin{aligned}
\lim _{t_{b} \rightarrow \infty} \pi \frac{d C_{\mathrm{A}}}{d t_{b}}=&P\Delta V+\frac{2 \tilde{A} (\varrho -1) \left[a^2 (\varrho +1)-\ell (\ell \varrho +\ell-2 m \varpi)\right]}{\alpha K \ell \varpi^2}\\&+\mathcal{O}\left(\tilde{A}^3\right)\\ \sim &P\Delta V+ \frac{2 \tilde{A}(\ell^2-a^2)}{K\ell \alpha \varpi^2}.
\end{aligned}
\end{equation}
In the vanishing acceleration limit $\tilde{A}\to 0$, the result reduces to the one obtained in \cite{AlBalushi:2020ely,AlBalushi:2020rqe}, which means that the rate is proportional to the volume difference. Remarkably, it can be seen that the modification from the non-vanishing acceleration on the late time CA growth rate beyond the volume term can be quite large in the condition of $\varpi^2 \ll\tilde{A}$, which means that the rate is not proportional to the volume difference anymore. It signifies that in a parameter regime
\begin{equation}\label{smcda}
\frac{\ell}{r_+}\ll A r_+ < A r\ll 1,
\end{equation}
the growth rate of the CA complexity at late times for the sufficiently large and slowly accelerating Kerr-AdS black hole is not controlled by the volume difference, but instead, gets non-ignorable drift from the conical singularities on the poles.

\section{CV complexity of formation}\label{cvfor}
Now we study the complexity of formation with the CV proposal, cf. Fig. \ref{pic0}. By symmetry the regularized volume of the $t=0$ extremal codimension-one bulk slice surface for the accelerating black hole is
\begin{equation}
\begin{aligned}
&C_{\mathrm{\mathcal{V}}}(t_b=0)=\max _{\partial \mathcal{B}=\Sigma}\left[\frac{\mathcal{V}(\mathcal{B})}{G_{N} \ell_{\mathrm{bulk}}}\right]\\&=\frac{4 \pi}{G_{N} \ell} \int_{0}^{\pi} d \theta \sin \theta \int_{r_{+}}^{r_{\max }} d r \frac{\rho}{K\Upsilon^3} \sqrt{\frac{\left(r^{2}+a^{2}\right)^{2}}{\Delta_r}-\frac{a^{2} \sin ^{2} \theta}{\Delta_{\theta}}},
\end{aligned}
\end{equation}
which can be obtained by extremalizing the volume functional.
Then the complexity of formation can be achieved by evaluating it and subtracting from it the analogous result of the vacuum,
\begin{equation}\label{cvf}
\Delta C_{\mathrm{\mathcal{V}}}(t_b=0)=C_{\mathrm{\mathcal{V}}}(t_b=0)-C_{\mathrm{\mathcal{V}}}(t_b=0, m=0).
\end{equation}

To calculate the complexity of formation, we must work out the IR bulk cutoff surface $r_{\text{max}}$ \cite{Anabalon:2018qfv}. We should find a coordinate transformation between the metric (\ref{met}) and the Fefferman-Graham coordinates
\begin{equation}
d s^{2}=\frac{\ell^{2}}{z^{2}} d z^{2}+z^{-2}\left(\gamma_{a b}^{(0)}+z^{2} \gamma_{a b}^{(2)}+\ldots\right) d x^{a} d x^{b}
\end{equation}
by series expansion of the coordinates $r$ and $\theta$ in $z\to 0$ as
\begin{align}
y=\frac{1}{r}=-A x-\sum_{n=1}^{\infty} \mathbb{F}_{n}(x) z^{n},\;
\cos \theta = x+\sum_{n=1}^{\infty} \mathbb{G}_{n}(x) z^{n}.\label{coordt2}
\end{align}

\begin{figure*}[htb!]
\begin{center}
\includegraphics[width=6in,angle=0]{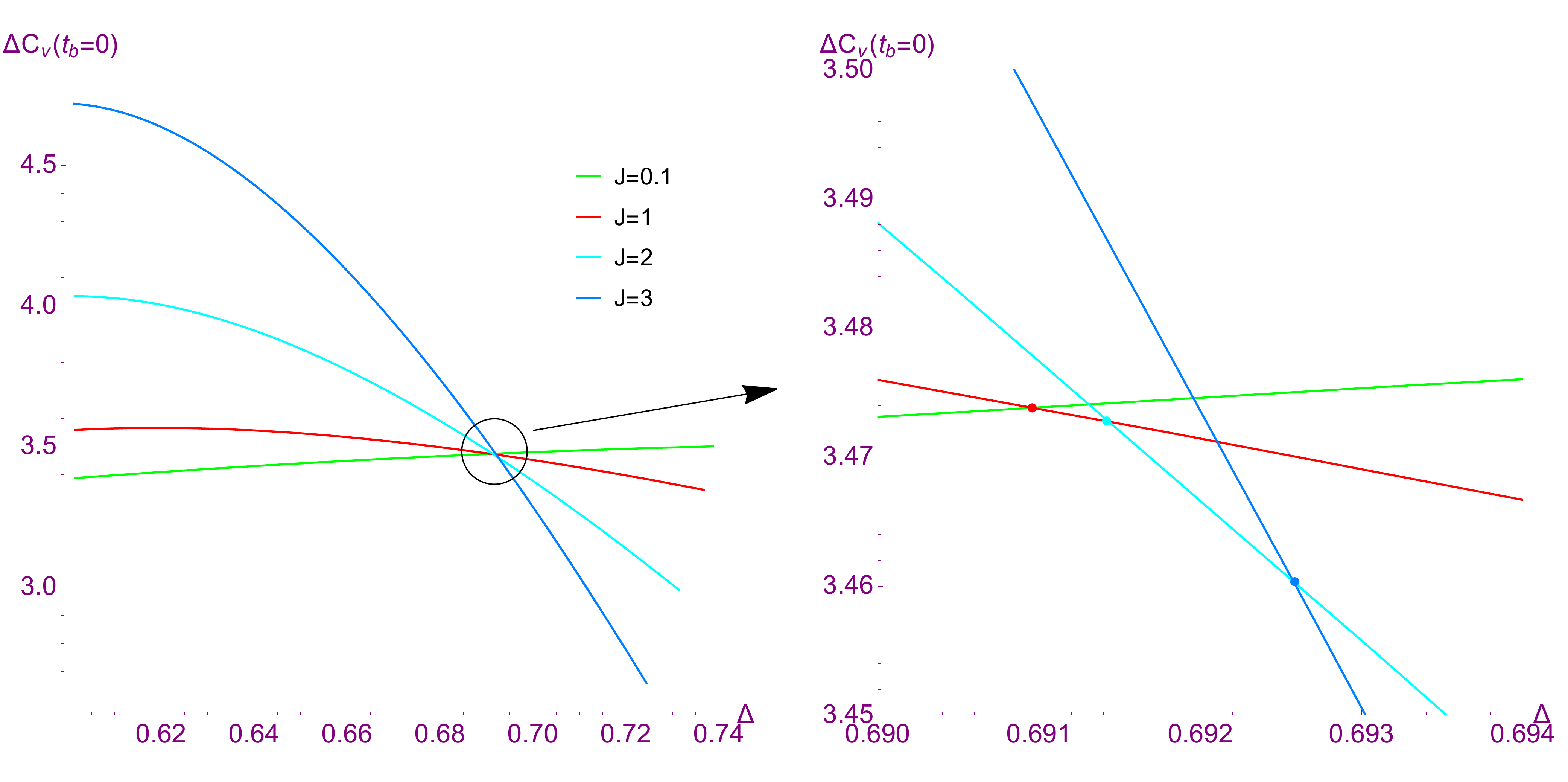}
\end{center}
\vspace{-5mm}
 \caption {Variations of the complexity of formation with respect to the average conical deficit with $S=15,\,\mu_-=1/10\,,P=3/(8\pi)$.}\label{pic2}
\end{figure*}

\begin{figure*}[htb!]
\begin{center}
\includegraphics[width=6in,angle=0]{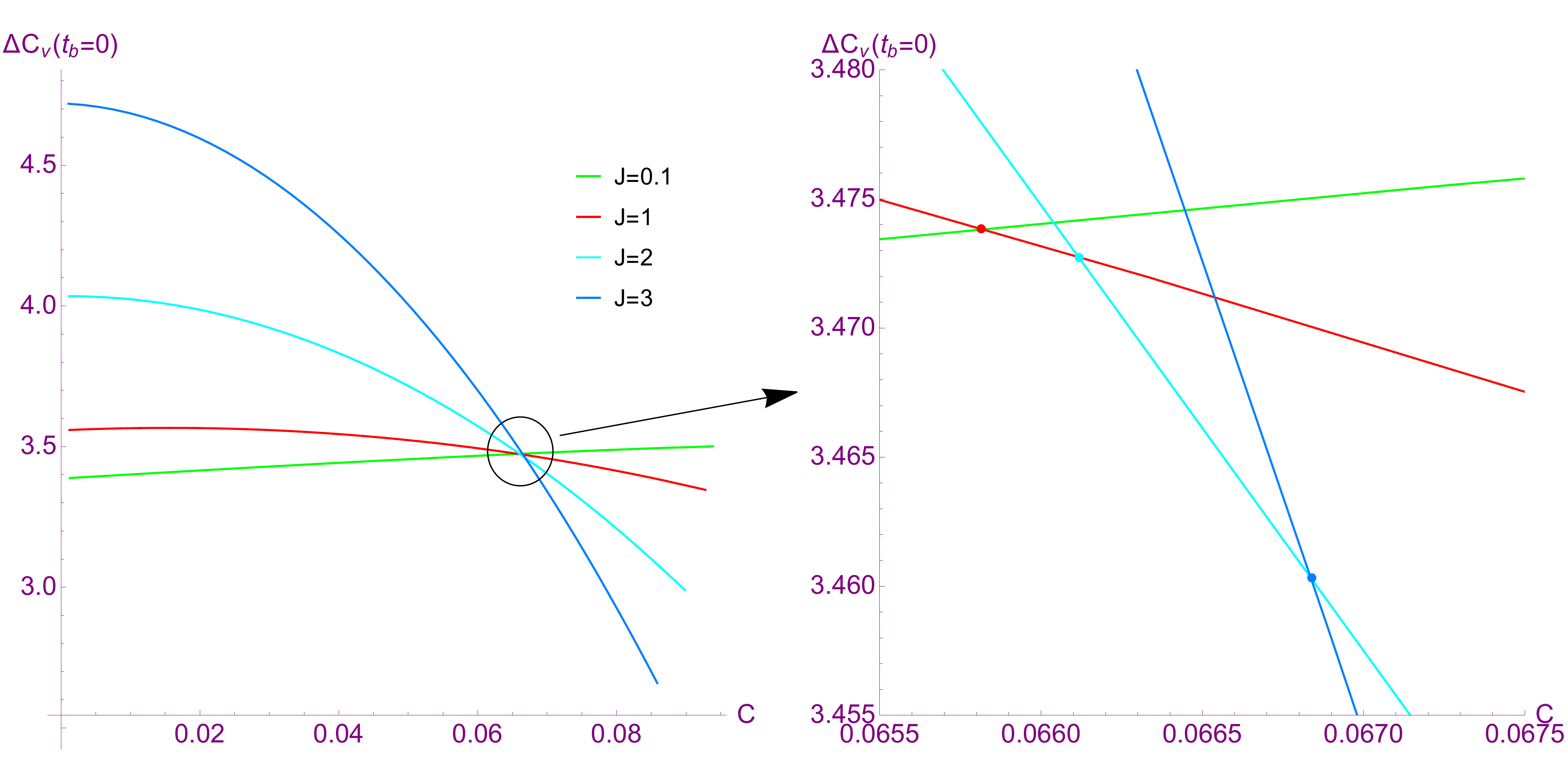}

\end{center}
\vspace{-5mm}
 \caption {Variations of the complexity of formation with respect to the differential conical deficit with $S=15\,,\mu_-=1/10\,,P=3/(8\pi)$.}\label{pic1}
\end{figure*}

Upon requiring $g_{zz}=\ell^2/z^2$ and $g_{za}=0$ at each order of $z$, we can get all $\mathbb{F}_n$ and $\mathbb{G}_n$ besides the conformal function $\mathbb{F}_1$, for which we choose $\mathbb{F}_{1}(x)=-\Upsilon(x)^{3}/\alpha \omega(x)$ \cite{Anabalon:2018qfv}. For instance, $g_{zx}=0$ gives
\begin{equation}
\begin{aligned}
&\frac{1}{\ell^{2}}\sum n G_{n} z^{n-1} \left(1+\sum G_{n}^{\prime}(x) z^{n}\right) \Upsilon_{a}^{2}\left(x+\sum \frac{F_{n}}{A} z^{n}\right) \\
=&-\sum n F_{n} z^{n-1}\left(A+\sum F_{n}^{\prime}(x) z^{n}\right) X\left(x+\sum G_{n} z^{n}\right),
\end{aligned}
\end{equation}
where
\begin{align*}
X(\xi) &=\left(1-\xi^{2}\right)\left(1+2 m A \xi+a^{2} A^{2} \xi^{2}-a^{2} \xi^{2} / \ell^{2}\right), \\
\Upsilon_{a}(\xi) &=\sqrt{1+a^{2} A^{2} \xi^{4}-A^{2} \ell^{2} X(\xi)}.
\end{align*}
Specifically, at the leading order $\mathcal{O}\left(z^{-2}\right)$, we have $G_{1}(x)=-A \ell^{2} F_{1}(x) X(x)/\Upsilon_{a}^{2}(x)$.
After tedious calculation, being sufficient to truncate the series (\ref{coordt2}) at $n=4$,  we can obtain the boundary metric
\begin{equation}
\begin{aligned}
d s_{(0)}^{2}&=\frac{\Upsilon_{a}^{4}}{\Sigma_{F G}^{3} F_{1}^{2}}\left(X(x)\left[a A^{2} x^{2} \frac{d t}{\alpha}-\left(1+a^{2} A^{2} x^{2}\right) \frac{d \varphi}{K}\right]^{2}\right.\\&\quad\left.-\frac{\Upsilon_{a}^{2}}{\ell^{2}}\left[\frac{d t}{\alpha}-a\left(1-x^{2}\right) \frac{d \varphi}{K}\right]^{2}\right)+\frac{\Upsilon_{a}^{2}}{F_{1}^{2} X} d x^{2}.
\end{aligned}
\end{equation}

To further elucidate the conformal degree of freedom, we specify the function $\omega(x)$ as \cite{Anabalon:2018ydc}
\begin{equation}
\omega^{2}=\frac{1-A^{2} \ell^{2} X(x) }{\alpha^{2}},
\end{equation}
which can recover from the boundary metric obtained in \cite{Hubeny:2009kz}. Thus, we have $x\in [-1, 1]$ and the coordinate transformation Eq. (\ref{coordt2}) reduces to $\theta$-dependent UV cutoff $z=\delta$ in the Fefferman-Graham coordinates, 
\begin{align} 
 r_{\max}& =-A x+\left[1-A^2\ell^2X(x)\right]\delta+\mathcal{O}(\delta^2),\label{coordt3}\\
 \cos \theta &= x+\frac{A\ell^2X(x)}{\Upsilon_a^2}\left[1-A^2\ell^2X(x)\right]\delta+\mathcal{O}(\delta^2)\label{coordt4},
\end{align}
where $\delta$ can be viewed as a short-distance cutoff in the dual CFT. In the limit $\delta\to 0$, $r_{\max}$ and $r_{\max}^{m=0}$ become identical. Then we can numerically calculate the CV complexity of formation for the accelerating Kerr-AdS black hole by substituting Eqs. (\ref{coordt3}) and (\ref{coordt4}) into Eq. (\ref{cvf}).

To see how the behavior of the complexity of formation is affected by the acceleration of the black hole and facilitate thermodynamic ensemble selection, we employ a compact mass formulation for the black hole \cite{Gregory:2019dtq},
\begin{equation}
\begin{aligned}
\frac{4\pi M^{2}}{\Delta S}=\left(1+\frac{8 P S}{3 \Delta }+\frac{\pi Q^{2}}{\Delta S}\right)^{2}+\left(\frac{4 \pi^{2} J^{2}}{\Delta^{2} S^{2}}-\frac{3 \Delta C^{2}}{2 P S}\right)\left(\frac{8 P S}{3 \Delta }+1\right),
\end{aligned}
\end{equation}
and encode the acceleration $A$ into $\Delta$ and $C$, which are respectively average conical deficit and differential conical deficit defined by
\begin{equation}
\Delta \equiv 1-2\left(\mu_{+}+\mu_{-}\right),\;C \equiv \frac{\mu_{-}-\mu_{+}}{\Delta}. \label{del}
\end{equation}
Notice from Eq. (\ref{mupm}), we must take $0 \leq C \leq \operatorname{Min}\{1/2, (1-\Delta2)/ (2\Delta) \}$ \cite{Gregory:2019dtq}.

To be specific, we study how the average conical deficit and differential conical deficit affect the CV complexity of formation in the ensemble with fixed $S$, $J$, $\mu_-$, and $P$ [cf. Eq. (\ref{tfl})]. As shown in Figs. \ref{pic2} and \ref{pic1}, we find that when the angular momentum of the black hole is small, which corresponds to the static limit, the CV complexity of formations increases with both $\Delta$ and $C$. In reverse, when the angular momentum increases, typically in the extremal regime, the CV complexity of formations decreases with both $\Delta$ and $C$.

\section{Conclusions}

We explored the holographic complexity of the rotating quantum gravity TFD state dual to the slowly accelerating Kerr-AdS black hole. After implicitly obtaining the form of the WDW patch, we evaluated the growth rate of the CA complexity at late times and discovered a nontrivial contribution from the acceleration. Due to the conical singularities on the poles, it was shown that the growth rate no longer scales as volume difference in large black hole limit; especially, the effects of the acceleration can be quite prominent if the acceleration is not too small [cf. Eq. (\ref{smcda})].
 
Then we studied the CV complexity of formation for the black hole. Specifically, in the pressure-fixed ensemble, we encoded the acceleration of the black hole into the average conical deficit $\Delta$ and differential conical deficit $C$ of the two poles. It was shown that in the static limit of the black hole, the complexity of formation increases with both $\Delta$ and $C$, but in the extremal regime, the complexity of formation decreases with both $\Delta$ and $C$. 

From the astrophysical observation we know that rotating and accelerating black holes exist in our Universe. In the perspective of holography, accelerating and rotating black holes are dual to specific boundary TFD states. Our result extended the investigation of gravitational CA holographic complexity in \cite{Bernamonti:2021jyu} and showed a non-trivial contribution from the not-too-small acceleration of large rotating black hole to the TFD states that breaks the volume-scaling formulation proposed in \cite{AlBalushi:2020ely,AlBalushi:2020rqe}. Further analysis of the CV complexity of formation revealed that the dependence of the holographic complexity on the acceleration of the black hole behaves differently in the static limit and in the extremal limit. This exhibits a non-negligible interplay of the acceleration (or cosmic string tension) and the angular momentum on the TFD states. It is conceivable that the classical cosmic string, which affects the geometry of the black hole spacetime, can be studied in a boundary quantum system.

It is worth noting that due to a lack of symmetry, the causal structure of the four-dimensional accelerating and rotating Kerr-AdS black hole cannot be tractable analytically, further analysis of the CA and CV holographic complexity for all times cannot be carried out with existing technologies. Moreover, we just showed specific behavior of the CV complexity formation in an ensemble with invariant entropy, angular momentum, and pressure, albeit analogous analysis can be done in other ensembles.

\section*{Acknowledgements}
M. Z. is supported by the National Natural Science Foundation of China with Grant No. 12005080.  J. J.  is supported by the National Natural Science Foundation of China with Grant No. 12205014, the Guangdong Basic and Applied Research Foundation with Grant No. 217200003 and the Talents Introduction Foundation of Beijing Normal University with Grant No. 310432102.

\bibliography{refs}
\end{document}